\documentstyle[preprint,aps]{revtex}
\draft
\begin{document}
\newcommand{\eq}[1]{Eq.~(\ref{#1})}
\bibliographystyle{prsty}
\preprint{RUB-TPII-01/96 }
\title{Tensor Charges of the Nucleon \\ in the
SU(3) Chiral Quark Soliton Model}
\author{
Hyun-Chul Kim,
\footnote{Electronic address: kim@hadron.tp2.ruhr-uni-bochum.de}
Maxim V. Polyakov,
\footnote{Alexander von Humboldt Fellow,
on leave of absence from Petersburg Nuclear Physics
Institute, Gatchina, St. Petersburg 188350, Russia}
and Klaus Goeke
}
\address{
Institute for  Theoretical  Physics  II, \\  P.O. Box 102148,
Ruhr-University Bochum, \\
 D-44780 Bochum, Germany  \\       }
\date{April 1996}
\maketitle
\begin{abstract}
The tensor charges of the nucleon are calculated in the framework of
the SU(3) chiral quark soliton model.  The rotational $1/N_c$ and
strange quark mass corrections are taken into account
up to linear order.   We obtain the following numerical values of the
tensor charges: $\delta u=1.12$, $\delta d=-0.42$,
and $\delta s=-0.008$.  In contrast to the axial charges,
the tensor charges in our model are closer to
those of the nonrelativistic quark model, in particular,
the net number of the transversely polarized strange
quarks in a transversely polarized nucleon $\delta s$ is compatible
with zero.
\end{abstract}
\pacs{PACS:11.15.Pg, 12.40.-y, 13.88.+e, 14.20.Dh}

\noindent
{\bf 1.} There are three different twist-two
nucleon parton distributions $f_1 (x)$,
$g_1(x)$ and $h_1(x)$. The knowledge of them would provide us the complete
information about the leading-order hard processes. Two of these
distributions-- $f_1(x)$ and $g_1(x)$-- have been investigated in detail
theoretically and measured in deep-inelastic lepton scattering (for a
review see \cite{Kodaira}).
However, the third distribution $h_1(x)$ which is called
the transversity distribution is chirally odd,
so that it does not appear in inclusive deep-inelastic scattering
experiments.
It was discussed that $h_1(x)$ can be measured in the Drell-Yan
lepton-pair production~\cite{RalstonSoper,leptonpair}, direct photon
production, and heavy-quark production in polarized $pp$ collisions
\cite{Ji1} and in the pion production in deep inelastic
scattering~\cite{JaffeJi}. Recently  Bourrely and Soffer suggested that
$h_1(x)$ can be determined in the neutral gauge boson $Z$ production in
$pp$ collisions \cite{BourrelySoffer}.  The $h_1(x)$ is totally unknown
experimentally, while its measurement has been proposed by the RHIC
spin collaboration~\cite{RSC}, HERMES collaboration at
HERA~\cite{HERMES} and more recently by COMPASS collaboration at CERN
\cite{compass}.

Jaffe and Ji~\cite{JaffeJi1} demonstrated that the first moment
of $h_1(x)$ is related to the tensor charge of the nucleon:
\begin{equation}
\int_{0}^{1} dx \left(h_1 (x) - \bar{h}_1 (x)\right)\;=\;\delta q,
\end{equation}
where $\bar{h}_1 (x)$ is an antiquark transversity distribution,
$\bar{h}_1 (x)=-h_1 (-x)$.  The tensor charges
$\delta q$ are defined as the forward nucleon matrix element:
\begin{equation}
\langle N|\bar{\psi}_q \sigma_{\mu\nu} \psi_q | N\rangle\;=\;
\delta q  \bar{U}\sigma_{\mu\nu} U ,
\label{tenschdef}
\end{equation}
where $q$ denotes the flavour index ($q=u,d,s$) and $U(p)$ stands for
a Dirac spinor and $\sigma_{\mu\nu}=\frac{i}{2}[\gamma_\mu,
\gamma_\nu]$.
We introduce for convenience  a flavour--singlet and two octet
tensor charges:
\begin{eqnarray}
g^{(0)}_{\rm T} &=& \delta u + \delta d +\delta s, \\
g^{(3)}_{\rm T} &=& \delta u - \delta d , \\
g^{(8)}_{\rm T} &=& \frac{1}{\sqrt 3}(\delta u + \delta d -2 \delta s).
\end{eqnarray}

In contrast to the axial charges
the tensor ones depend on the renormalization scale already at one--loop
level. The corresponding anomalous dimension  has been
evaluated in Refs. \cite{Kodairaetal,Bukhvostovetal,ArtruMekhfi}:  $\gamma
= 2\alpha_s/3\pi$.  However, their dependence on
the normalization point is very weak:
 \begin{equation} \delta q (\mu^2)
\;=\;\left(\frac{\alpha_s (\mu^2)} {\alpha_s (\mu^2_{0})}
\right)^{\frac{4}{33-2 N_f}} \delta q (\mu^2_{0}).
\end{equation}
As $\mu\rightarrow\infty$ the $\delta q (\mu^2)$ is slowly vanishing.
This equation can be used to evolve the tensor charges from the low
normalization point (several hundreds MeV) pertinent to the
chiral quark-soliton model ($\chi$QSM) we are dealing with, to higher
normalization points.
Since the corresponding anomalous dimension is relatively small,
the value of the tensor charge at a higher normalization point is insensitive
to uncertainties of low normalization points relevant to
our model~\cite{KimPoGo}.

Quite recently, we have examined the tensor charges of the nucleon in
the framework of the SU(2) chiral quark-soliton model ($\chi$QSM)
and suggested the mechanism as to how the tensor charges are
different from the axial ones~\cite{KimPoGo}.
In the present paper, we extend the former investigation
to the case of three flavours.  This enables us to evaluate
the net number of the transversely polarized up, down and strange quarks
in a transversely polarized nucleon separately.

Since the tensor current is not related to any symmetry,
it can not be constructed as a Noether current.
 Hence, it is not obvious how to build up the tensor current
in the Skyrme model, because the corresponding Lagrangian
consists only of mesonic fields.  In contrast to the Skyrme model,
one can define unambiguously any quark current in the $\chi$QSM having
explicit quark degrees of freedom.

\noindent
{\bf 2.}
The $\chi$QSM is based on the interaction of dynamically massive
constituent quarks with pseudo-Goldstone meson fields.
It is characterized by the low-energy effective chiral lagrangian
given by the functional integral over
quark ($\psi$) in the background pion field
 \cite{DE,ATFR,DSW,DyPe1}:
\begin{equation}
\exp\left(iS_{\rm eff}[\pi(x)]\right)\; = \; \int {\cal D}\psi
{\cal D}\bar\psi
\exp{\left( \int d^4 x \bar\psi D \psi \right)},
\label{eq:Z}
\end{equation}
where $D$ is the Dirac operator
\begin{equation}
 D \; = \;  i \rlap{/}{\partial} -\hat{m} -MU^{\gamma_5}.
\label{Dirac_operator}
\end{equation}
$U^{\gamma_5}$ denotes the pseudoscalar chiral field
\begin{equation}
U^{\gamma_5}\;=\;\exp{i\pi^a \lambda^a \gamma_{5}}
\;=\;\frac{1+\gamma_5}{2}U\;+\;\frac{1-\gamma_5}{2} U^{\dagger}.
\end{equation}
$\hat{m}$ is the matrix of the current quark masses
$\hat{m}\;=\;\mbox{diag}(m_u, m_d, m_s)$ and $\lambda^a$ represent
the usual Gell-Mann matrices.
The $M$ stands for the dynamical quark mass arising as a
result of the spontaneous chiral symmetry breaking.

The effective chiral action given by \eq{eq:Z} is known to
contain automatically the Wess--Zumino term and the four-derivative
Gasser--Leutwyler terms, with correct coefficients. Therefore, at least the
first four terms of the gradient expansion of the effective chiral
lagrangian are correctly reproduced by \eq{eq:Z}, and chiral symmetry
arguments do not leave much freedom for further modifications.
\eq{eq:Z} has been derived from the instanton model of the QCD vacuum
\cite{DyPe1}, which provides a natural mechanism of chiral
symmetry breaking and enables one to express the dynamical mass $M$ and the
ultraviolet cutoff $\Lambda$ intrinsic in \eq{eq:Z} through the
$\Lambda_{QCD}$ parameter. It should be mentioned that \eq{eq:Z} is
of a general nature: one need not believe in instantons
and still use \eq{eq:Z}. The effective chiral theory \eq{eq:Z}
is valid for the values of the quark momenta up to the ultraviolet
cutoff $\Lambda$. Therefore, in using \eq{eq:Z} we imply that we are
computing the tensor charges at the normalization point about
$\Lambda\approx 600$~MeV.

An immediate application of the effective chiral theory \eq{eq:Z} is the
quark-soliton model of baryons \cite{DiPePo}.
According to these ideas the nucleon
can be viewed as a bound state of $N_c$ (=3) {\em valence}
quarks kept together
by a hedgehog-like pion field whose energy coincides by definition
with the aggregate energy of quarks from the negative Dirac sea.  Such a
semiclassical picture of the nucleon is justified in the limit
$N_c\rightarrow\infty$ -- in line with more general arguments by Witten
\cite{W}. Roughly speaking, the $\chi$QSM builds a bridge
between the naive valence quark model of baryons and the
Skyrme model.  The further studies showed that the $\chi$QSM is
successful in reproducing the static properties
and form factors of the baryons using just one parameter set and
adjusted in the mesonic sector to $m_\pi$, $f_\pi$ and $m_K$.
(see the recent review \cite{Review}).

The forward nucleon matrix element Eq.~(\ref{tenschdef}) in the
rest frame of the nucleon
is nonzero only for indices $\mu,\nu$ being space-like
$i,j=1,2,3$.
Using a $\gamma$-matrix property $\sigma_{\mu\nu}=
(i/2) \epsilon_{\mu\nu\alpha\beta} \sigma^{\alpha\beta}\gamma_5 $,
we can relate the operator in the left-hand side of
Eq.~(\ref{tenschdef}) in the rest
frame of the nucleon to $\bar{\psi}
\sigma_{0i} \gamma_5 \lambda^a \psi$.
 Hence, the tensor charges can be calculated as a nucleon forward matrix
element of  $\bar{\psi}\gamma_0 \gamma_5 \gamma_k \lambda^a\psi$.  It is
interesting to notice that the only difference between the axial and tensor
charges is the $\gamma_0$ matrix.  It implies that in the nonrelativistic
quark model (NRQM) the tensor charges coincide with the axial
ones~\cite{JaffeJi1,HeJi}:
\begin{eqnarray}
\nonumber
\delta u&=&\Delta u =\frac43, \\
\delta d &=& \Delta d=-\frac13 , \\
\nonumber
\delta s&=& \Delta s=0.
\label{nrres}
\end{eqnarray}
The tensor charges of the nucleon  can be related to the
following correlation function in Euclidean space:

\begin{equation}
-i \langle 0 | J_N (\vec{y}, \frac{T}{2}) \psi^\dagger
\gamma_0\gamma_k \lambda^a  \psi J^\dagger_{N}
(\vec{x}, -\frac{T}{2}) |0 \rangle
\label{corrf}
\end{equation}
at large Euclidean time $T$.  The nucleon current $J_N$ is built of
$N_c$ quark fields:
\begin{equation}
J_N(x)\;=\; \frac{1}{N_c !} \epsilon_{i_1 \cdots i_{N_c}}
\Gamma^{\alpha_1 \cdots
\alpha_{N_c}}_{JJ_3TT_3Y}\psi_{\alpha_1i_1}(x)
\cdots \psi_{\alpha_{N_c}i_{N_c}}(x).
\end{equation}
$\alpha_1 \cdots\alpha_{N_c}$ denote spin--flavour indices, while
$i_1 \cdots i_{N_c}$ designate colour indices.  The matrices
$\Gamma^{\alpha_1 \cdots\alpha_{N_c}}_{JJ_3TT_3Y}$ are taken to endow
the corresponding current with the quantum numbers $JJ_3TT_3Y$.

The nucleon matrix element of
$\bar{\psi}\gamma_0 \gamma_5 \gamma_k \lambda^a\psi$
can be computed as the Euclidean functional integral in the $\chi$QSM
\begin{eqnarray}
\langle N |\psi^\dagger \gamma_0\gamma_5\gamma_k
\lambda^a \psi | N \rangle & = &
\frac{1}{\cal Z} \lim_{T \rightarrow \infty} \exp{(ip_0 \frac{T}{2}
- ip'_{0} \frac{T}{2})} \nonumber \\
& \times & \int d^3 x d^3 y
\exp{(-i \vec{p'} \cdot \vec{y} + i \vec{p} \cdot \vec{x})}
\int {\cal D}U \int {\cal D} \psi \int {\cal D}\psi^\dagger
\nonumber \\
& \times & \; J_{N}(\vec{y},T/2)\psi^\dagger \gamma_0
\gamma_5 \gamma_k \lambda^a \psi J^{\dagger}_{N} (\vec{x}, -T/2)
\nonumber \\ & \times &
\exp{\left[\int d^4 z \psi^\dagger D \psi \right ]}.
\label{Eq:ev}
\end{eqnarray}

In the large $N_c$ limit the integral over Goldstone fields
$U$ can be calculated by the steepest descent method (semiclassical
approximation).  The corresponding
saddle point equation admits a static soliton solution, an example of
which is the hedgehog field configuration:
\begin{equation}
U_s(\vec{x})\;=\; \left ( \begin{array}{cc}
U_0 & 0 \\ 0 & 1 \end{array} \right ),
\label{Eq:embed}
\end{equation}
where $U_0$ is the SU(2) chiral matrix of the form:
\begin{equation}
U_0\;=\;\exp{[i\vec{n}\cdot\vec{\tau}P(r)]} .
\end{equation}
The $P(r)$ denotes the profile function satisfying the boundary condition
$P(0)=\pi$ and $P(\infty)=0$, which is determined by solving the
saddle point equations (for details see Ref.~\cite{Review}).
The soliton is quantized by introducing collective coordinates
corresponding to $SU(3)_{fl}$ rotations of the soliton in flavour space
(and simultaneously $SU(2)_{spin}$ in spin space):
\begin{equation}
U(t,\vec{x})=R(t)U_s(\vec{x})R^\dagger(t),
\end{equation}
where $R(t)$ is a time--dependent $SU(3)$ matrix.
The quantum states arising
from this quantization have the quantum numbers of baryons.
In the large $N_c$ limit the soliton angular velocity
$\Omega=R^\dagger(t)\dot{R}(t)$ is
parametrically small, so that we can use the
angular velocity as a small parameter. Recently, it was demonstrated
\cite{WakamatsuWatabe,Chetal} that taking into account the first order
rotational corrections one can solve old problems of underestimate of the
nucleon axial constants and magnetic moments in the chiral soliton
model of the nucleon.
 Also it is worth noting that the correct non-relativistic
quark model results for axial and tensor charges Eq.~(\ref{nrres}) can be
obtained in the non-relativistic limit of the $\chi$QSM only
if the first order
rotational corrections are considered \cite{PraBloGo}.
The next source of the corrections to the leading order result is the
effects of $SU(3)$ symmetry violation caused
by the nonzero strange quark mass.
We calculate the $SU(3)$ symmetry breaking corrections linear in $m_s$.

We follow closely the formalism described in Ref.~\cite{Review}
and hence we present below only the results without any detail
(they will be given elsewhere). The tensor charges of the nucleon
have the following structure (order of each term in $1/N_c$ and $m_s$ is
shown explicitly):

\begin{eqnarray}
g^{(a)}_{\rm T}&=& N_c\left\{ {\cal T}_0
\langle D^{(8)}_{a3} \rangle_N + \frac{{\cal T}_1}{N_c}
\langle D^{(8)}_{a3} \rangle_N
+ \frac{{\cal T}_2}{N_c}  \langle d_{3pq} D^{(8)}_{ap} J_q \rangle_N
+ \frac{{\cal T}_3}{N_c}  \langle D^{(8)}_{a8} J_3 \rangle_N
+  m_s  {\cal T}_4 \langle D^{(8)}_{88} D^{(8)}_{a3}\rangle_N\right.
 \nonumber \\
&+&  \left. m_s  {\cal T}_5 \langle D^{(8)}_{83}D^{(8)}_{a8}\rangle_N
+  m_s  {\cal T}_6 \langle d_{3pq}D^{(8)}_{8p} D^{(8)}_{aq}\rangle_N
+ O(\frac{1}{N_c^2}) + O(\frac{m_s}{N_c}) + O(m_s^2) \right\}
\label{Eq:gt38}
\end{eqnarray}
for $a = 3, 8$ and
\begin{equation}
g^{(0)}_{\rm T}= \sqrt 3 {\cal T}_3
\langle J_3 \rangle_N
+\sqrt 3 m_s N_c {\cal T}_5\langle D^{(8)}_{83}\rangle_N,
\label{Eq:gt0}
\end{equation}
where $\langle O \rangle_N$ denotes the average over rotational
state of the quantized soliton corresponding to the nucleon,
$J_a$ ($a=1,\ldots,8$) is the operator of infinitesimal $SU(3)$ rotation,
for $a=1,2,3$ it coincides with the operator of angular momentum.
The quantities ${\cal T}_i$ in Eqs.(\ref{Eq:gt38},
\ref{Eq:gt0}) can be calculated as functional traces of the form:

\begin{equation}
{\cal T}_i =\mbox{Sp}(\frac{1}{D(U_s)}
\Gamma_1^i \frac{1}{D(U_s)} \Gamma_2^i),
\end{equation}
where $D(U_s)$ is a Dirac operator Eq.~(\ref{Dirac_operator}) in the static
chiral soliton field Eq.~(\ref{Eq:embed}), $\Gamma^i_{1,2}$ are operators
which are local in coordinate space and
generically non-local in time. The explicit expressions
for ${\cal T}_i$ will appear elsewhere.

In order to evaluate Eqs. (\ref{Eq:gt38}, \ref{Eq:gt0})
numerically, we employ the Kahana-Ripka discretized basis
method~\cite{KaRi,Review}.
The constituent quark mass is fixed to 420 MeV
in our model by producing best the SU(3) baryon mass
splittings~\cite{Blotzetal}.
All other relevant static baryon observables and
form factors are also well reproduced in the model
\cite{Review} for this constituent quark mass.
To make sure of the numerical calculation,
we compare our results for ${\cal T}_i$ with the analytical
ones of the gradient expansion justified in the limit of
large soliton size.  Our numerical procedure reproduces
within few percent the analytical results of the
gradient expansion  for each ${\cal T}_i$ separately
in large soliton size limit.

The results of our calculations are summarized in Tables~I--II.
We see that the rotational $1/N_c$ corrections are of great importance
numerically, whereas the $SU(3)$ symmetry breaking corrections are
relatively small. Unlike the axial
charges \cite{BlPrGo,BlPoGo}, the tensor ones in our model are closer to
their values in nonrelativistic quark model,
in particular the strangeness contribution to the tensor charge
$\delta s$ is compatible with zero, while
the analogous contribution to the axial charge $\Delta s$ in the same
model and in the experiment is negative and distinctive from
zero \cite{BlPrGo}.

\noindent
{\bf 3.} To summarize, we investigate the tensor charges in
the SU(3) chiral quark-soliton model which is also called the
semibosonized SU(3) Nambu-Jona-Lasinio model.  For the first time,
the octet tensor charge $g^{(8)}_{\rm T}$ and hence the net number of the
transversely polarized strange quarks in a transversely polarized nucleon
$\delta s$ are calculated.
 An interesting feature of our model is that it
predicts the negative nonzero number of the  polarized strange
quarks $\Delta s$ in the longitudinally polarized nucleon
\cite{BlPrGo,BlPoGo}, which is consistent with the corresponding
experimental value, whereas it yields the number of
the transversely polarized strange quarks $\delta s$
in a transversely polarized nucleon compatible with zero.

The dynamical origin of the difference between the axial and tensor
charges in our model can be related to the qualitatively different
behaviour of the charges with soliton size \cite{KimPoGo}.
The detailed discussion of this issue
will be published elsewhere.

This work has partly been supported by the BMBF, the DFG
and the COSY--Project (J\" ulich).  The work of M.P. is supported
by the Alexander von Humboldt Foundation.
\begin{table}
\caption{Tensor charges $g^{(0)}_{\rm T}$, $g^{(3)}_{\rm T}$ and
$g^{(8)}_{\rm T}$ with the constituent quark mass $M=420\;\mbox{MeV}$.
The current quark mass $m_s$ is chosen as $m_s = 180$ MeV.
 The final model predictions are given by
${\cal O}(\Omega^1, m^1_{s})$.}
\begin{tabular}{cccc}
   & ${\cal O}(\Omega^0, m^0_{s})$ &  ${\cal O}(\Omega^1, m^0_{s}) $ &
${\cal O}(\Omega^1, m^1_{s})$ \\
\hline
$g^{(0)}_{\rm T}$     &   0  & 0.69 & 0.70 \\
$g^{(3)}_{\rm T}$     &   0.79  & 1.48 & 1.54 \\
$g^{(8)}_{\rm T}$     &   0.09  & 0.48 & 0.42 \\
\end{tabular}
\caption{Each flavour contribution to the tensor charges
as varying the constituent quark mass $M$.
The current quark mass $m_s$ is chosen as $m_s = 180$ MeV.}
\begin{tabular}{cccc}
$M$~[MeV]   & $\delta u$ &  $\delta d$  &  $\delta s$ \\
\hline
     370 &   1.18  & -0.41 &  0.002 \\
     400 &   1.14  & -0.41 & -0.004 \\
     420 &   1.12  & -0.42 & -0.008 \\
     450 &   1.12  & -0.41 & -0.02  \\
\end{tabular}
\end{table}

\vfill\eject

\vfill\eject
\newpage
\pagebreak
\end{document}